\documentclass[twocolumn,showpacs,preprintnumbers,amsmath,amssymb]{revtex4}
\usepackage{graphicx}
\usepackage{dcolumn}
\usepackage{bm}

\begin{document}

\title{Stability in the evolution of random networks}

\author{L.P. Chi}
\email{chilp@iopp.ccnu.edu.cn} \affiliation{Institute of Particle
Physics, Hua-Zhong (Central China) Normal University, Wuhan
430079, P.R. China}
\author{C.B. Yang}
\affiliation{Institute of Particle Physics, Hua-Zhong (Central
China) Normal University, Wuhan 430079, P.R. China}
\author{X. Cai}
\affiliation{Institute of Particle Physics, Hua-Zhong (Central
China) Normal University, Wuhan 430079, P.R. China}

\date{\today}

\begin{abstract}

With a simple model, we study the evolution of random networks
under attack and reconstruction. We introduce a new quality,
\textit{invulnerability} $I(s)$, to describe the stability of the
system. We find that the network can evolve to a stationary state.
The stationary value $I_{c}$ has a power-law dependence on the
initial average degree $\langle k \rangle$, with the slope is
about $-1.485$. In the stationary state, the degree distribution
is a normal distribution, rather than a typical Poisson
distribution for general random graphs. The clustering coefficient
in the stationary state is much larger than that in the initial
state. The stability of the network depends only on the initial
average degree $\langle k \rangle$, which increases rapidly with
the decrease of $\langle k \rangle$.

\end{abstract}


\maketitle

\section{Introduction}

Many systems can be represented by networks, a set of nodes joined
together by links indicating interactions. Social
networks~\cite{wasserman}, the Internet~\cite{BA1}, food
webs~\cite{williams}, transportation
networks~\cite{Li,Chi,Latora}, and linguistic
networks~\cite{Cancho} are just some examples of such systems. The
investigation of complex networks was initiated by Erd\H{o}s and
R\'{e}nyi~\cite{ER1}. They proposed and studied one of the
earliest theoretical models of a network, the random graph. In a
random graph, $N$ labelled nodes are connected by $n$ edges, which
are chosen randomly from the $N(N-1)/2$ possible edges. It is
trivial to show that the connection probability is
$p=n/[N(N-1)/2]$. The number $k$ of edges connecting one node to
others is called the degree of that node. The average degree of
the graph is $\langle k \rangle=2n/N=p(N-1)\simeq pN$ if $p\ll 1$.
The degree distribution for a random network is given by a
Poissonian distribution.

Recently the increasing accessibility of databases of real
networks and the availability of powerful computers have made
possible a series of empirical studies on complex networks. Thus,
other two main streams of topics were proposed and investigated in
depth. One is the small-world networks introduced by Watts and
Strogatz~\cite{WS1,WS2}. Such networks are highly clustered like
regular lattices, yet have small characteristic path lengths like
random graphs. The other is the scale-free networks proposed by
Barab\'{a}si and Albert~\cite{BA2,BA3,BA4}, based on two generic
mechanisms, growth and preferential attachment. Those networks
have scale-free power-law degree distributions.

With increased threats of hacker attacks and routers malfunction,
etc., research in the field of network robustness has attracted
much attention~\cite{attack1}. Albert and his collaborators have
shown that scale-free networks, at variance with random graphs and
small-world networks, are almost unaffected by errors while
vulnerable to attacks. That is, the ability of their nodes to
communicate is almost unaffected by the failure of some randomly
chosen nodes, but the removal of a few most connected nodes can
damage the networks. Recently the network efficiency of errors and
attacks on scale-free networks has also been
studied~\cite{efficiency}. In those studies, the damaged nodes are
removed from the network. Thus the size of the network decreases
with the evolution of the system.

Based on the physics of network tolerance, using a simple
model on evolving network, we are trying to study the robustness
of a dynamical evolving network. We will consider the reconstruction
of the links of the damaged nodes instead removing the nodes from the network.
Thus the size of the system remains
unchanged. This is more closer to the evolution of real networks.
The paper is organized as follows. The evolution model is presented in next section.
Section III is devoted to analyze the results of the model, along
with a definition of how we describe the stability of the network.
In last section we conclude and give a brief discussion.

\section{An Evolution Model}

Despite the fine work of studies on network tolerance, little
effort has been made on the reconstruction of the attacked
network. The aim of our model is to investigate the stability of
the network during the evolution in terms of attacks and
reconstructions. In other words, the nodes damaged will not be
removed from the network, instead they will be reconnected in
certain way. Assume that all information on the former links of
the damaged nodes has been lost, the damaged nodes have to be
connected randomly to other nodes in the network again. In this
way, we keep the size of the system constant. We do not try to
consider a case with increasing size in the evolution, since the
process of system's size increase can be much slower than the
frequently happening damage and reconstruction. We try to
investigate the effect of such a reconstruction on the evolution
of the network. For this purpose, we first setup a random netweok
with $N$ nodes and connection probability $p$, then let the
network evolve according to the following rules: (i) Find the node
with the highest degree $k_{max}$. Since this node has highest
connections to other nodes, it is most likely attacked and is the
most vulnerable site in the network. For simplicity, we assume
that only a node with highest connections suffers attack and to be
reconnected. If several nodes happen to have the same highest
degree of connection, only one (randomly chosen) of them is
assumed to be damaged in the attack. (ii) Reconnect this node with
the other nodes in the network with the reconstruction probability
$p_{re}=p$. Steps (i) and (ii) are repeated to a prefixed number
of times. This number should be large enough to enable the system
to reach (possibly existing) stationary state. In our simulation,
it is chosen to be 10 million. The reconnection process represents
the effort of reconstructing the network. We set the reconnection
probability $p_{re}$ the same as the initial construction
probability $p$ to reduce the number of parameters in the model.
This set also implies that the information does not increase from
initial construction to later reconnection. Thus, the evolution of
the network is in fact a process of bing damaged and subsequent
reconstructing, similar to the evolution of real networks.

The interest in our model investigation is to find out whether
there exists a stationary state in the evolution of the network,
and if it exists, what are the properties of the stationary state
and their dependences on the parameters in the model.

\section{Simulation Results}

\begin{figure}
\includegraphics{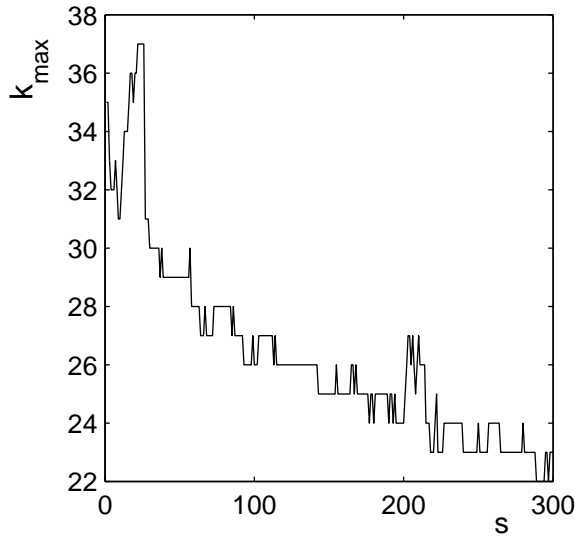}
\caption{\label{fig1}Plots of $k_{max}$ versus step $s$ with
$N=1000$, $p_{re}=p=0.02$.}
\end{figure}

The model defined in Sect. II has several nontrivial consequences.
It can be easily seen that the network has a tendency to decrease
the maximum number of connections among the nodes at a long time
scale, because the nodes with highest connections will be damaged
and reconnected randomly. From  the evolution rules, the total
number of connections in the network will also decrease generally
before a stationary state is reached.

To get some idea on the safety of the network, we can have a look
at the behavior of the maximum degree of connection of the
network. In Fig. 1, we give a snapshot of the maximum degree
$k_{max}$ versus time step $s$ for a network with $N=1000$ nodes
and connection probability $p_{re}=p=0.02$. Apart from some
fluctuations, it decreases in the evolution. From intuition, a
node with less links to others will be attacked less frequently.
Thus a network with smaller maximum connection degree is safer. To
describe the safety of the network, we introduce a new quantity,
the invulnerability $I(s)$, which is analogous in definition to
the \textit{gap} $G(s)$ in the Bak-Sneppen model~\cite{BS1,BS2}.
Considering an evolution of network with maximum degrees $k_{\rm
max}(1), k_{\rm max}(2), \cdots, k_{\rm max}(s)$, invulnerability
$I(s)$ at time $s$ is defined
\begin{equation}
I(s)=1/Min\{k_{\rm max}(i)\} \ \ \textbf{for\ \ \ } i\leq s\ ,
\end{equation}
i.e., the inverse of the minimum of all the maximum degrees in the
evolution before moment $s$. Initial value of $I(s)$ is equal to
$1/k_{max}(1)$. $I(s)$ reflects the attack tolerance of the
network. When $I(s)$ is small, the network is vulnerable to
attack. Obviously from definition, $I(s)$ is a non-decreasing
function of evolution, and some fluctuations in $k_{\rm max}$ have
been filtered out.

\begin{figure}
\includegraphics{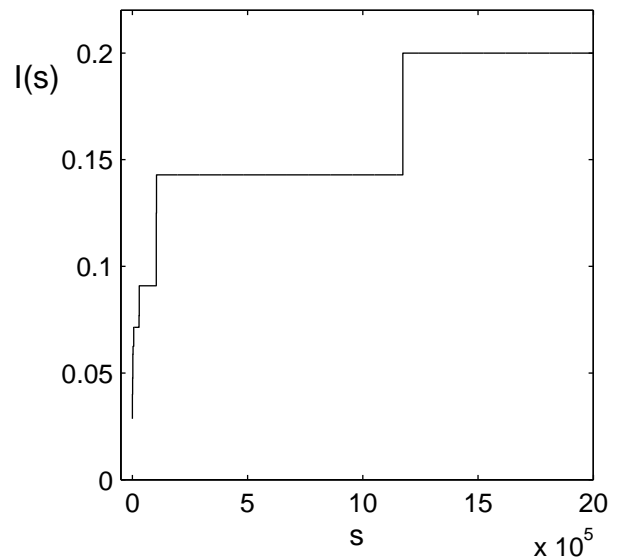}
\caption{\label{fig2}Plots of invulnerability $I(s)$ versus $s$
for network with $N=1000$ nodes and connection probability
$p_{re}=p=0.02$.}
\end{figure}

\begin{figure}
\includegraphics{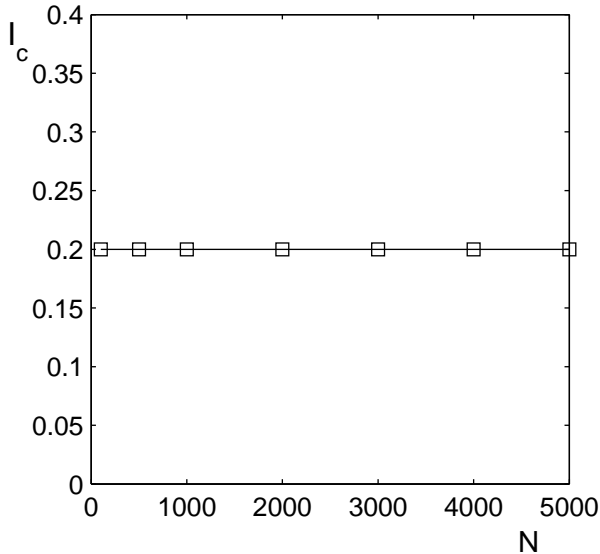}
\caption{\label{fig3}The stationary value of $I_{c}$ as a function
of network size $N$ when $\langle k \rangle =20$ fixed.}
\end{figure}

Fig. 2 shows $I(s)$ versus step $s$ with network size $N=1000$ and
probability $p_{re}=p=0.02$. We observe that $I(s)$ increases very
quickly at small $s$ but slowly at large $s$ and finally reaches a
constant value when $s$ is large enough. The increase of $I(s)$
indicates that the system is getting more and more safe in general
by experiencing continuous attacks and reconstructions. This
figure is similar to the envelope function of Bak-Sneppen
evolution model. Without interference from outside world, the
network evolves to a stationary state. And the process takes place
over a very long transient period. In Fig. 3, we present the
stationary value $I_{c}$ as a function of the network size $N$
under a fixed initial average degree $\langle k \rangle=20$. We
find $I_{c}$ stays unchanged at 0.2 with $N$ shifted from 100 to
5000. This result is interesting, because it shows that the
stationary value $I_{c}$ depends not on the network size $N$ and
probability $p_{re}=p$ separately but through the initial average
degree $\langle k\rangle=pN$. Such a dependence can be expected
considering the facts that the initial degree distribution has an
average $\langle k\rangle=pN$ with variance $\sigma^2=\langle
k\rangle=pN$ and that the average number of links to the
reconstructed nodes is also $\langle k\rangle=pN$. As a result,
the properties of the evolution are mainly determined by the value
of initial average degree $\langle k\rangle=pN$. To get the
relationship between $I_c$ and $\langle k\rangle$ we show in Fig.
4 $I_{c}$ as a function of $\langle k \rangle$ in a log-log plot.
We find that $I_{c}$ has a power-law dependence on the average
degree $\langle k \rangle$,
\begin{equation}
I_{c}(\langle k \rangle)\propto{\langle k \rangle}^{-\tau},
\label{power}
\end{equation}
\noindent where the exponent $\tau$ is about 1.485. Fig. 4
illustrates that after the network has relaxed to the stationary
state, the stability of the network will increase rapidly with the
decrease of average degree $\langle k \rangle$ in the initial
state. Thus, when the initial average degree $\langle k \rangle$
is small, i.e., less communications and interactions in the
network, the system will be more stable. We would like to mention
that the value of $\tau=1.485$ results from the nonlinear
interactions among the nodes.

\begin{figure}
\includegraphics{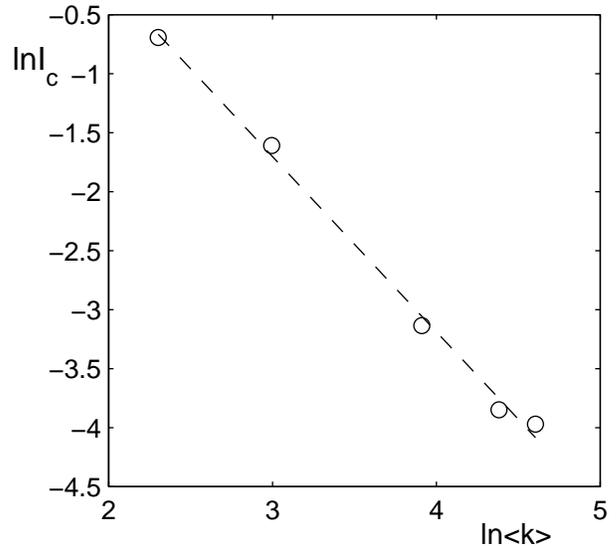}
\caption{\label{fig4}$\ln I_{c}$ versus $\ln \langle k \rangle$. A
linear fit is shown as dash line with slope $-1.485$.}
\end{figure}

In order to offer a further information of the network in the
stationary state, one can compare some structural properties of the
network in the initial state with those in the
stationary state.

Because of the reconnection of a damaged node to other randomly
chosen ones, the nodes with modest degree of connection may
increase their links. As a result of the reconstruction, the
distribution of connections $k$ changes in the evolution of the
system. To see how dramatically the change is, we compare the
distributions at initial state and the state in which $I(s)$
remains unchanged. For this purpose, we construct a random graph
with $N=10000$ nodes and $p=0.005$ and plot its degree
distribution in the initial state by circles in Fig. 5(a). Because
of the randomness of the initial connections, the initial degree
distribution of a random graph is a Poisson distribution for large
$N$~\cite{BA4},
\begin{equation}
P(k)=e^{-\langle k \rangle}\langle k \rangle^{k}/ k! \,
\end{equation}
with peak at about $\langle k \rangle=50$. This theoretical
expectation is shown in the figure as the solid line. Then let the
network evolve with the given simulation rules. When the system
reaches the stationary state, we draw its degree distribution, as
represented by squares in Fig. 5(a). We fit the distribution to a
normal distribution, and find that the average degree is $\langle
k \rangle =17$ and standard deviation is $\sigma =7$. One can see
that the mean degree of connection decreases by a factor 3 in the
evolution. One may notice that a Poissonian distribution can be
approximated by a normal distribution when the mean value $\langle
k\rangle$ is large. For Poissonian distribution, the width of the
distribution is determined also by the mean value $\langle
k\rangle$ as $\sigma=\sqrt{\langle k\rangle}$. In a normal
distribution, the width has no correlation with its mean value.
Therefore, the stationary degree distribution can not be described
by a Poissonian distribution. To investigate the dependence of
such a change on the initial $\langle k \rangle$, we do the same
investigation for a network with $N=10000$ nodes and $p=0.015$.
This time, the initial $\langle k \rangle$ is 150. The initial and
final state distributions of connection are shown in Fig. 5(b),
represented by circles and squares, respectively. Both
distributions can be fitted very well by Gaussian distributions,
with $\langle k \rangle =150$ for the initial distribution as
expected and 77 for the final state. The width of the final state
connection distribution is now $\sigma =18.5$. From these two
evolutions one can conclude that the change of connection
distributions depends on the initial mean degree nontrivially. The
higher the initial mean degree, the wider of the distribution for
the stationary state. In fact, the two average degrees above in
the stationary states satisfy the relation in Eq. (\ref{power})
with the same $\tau$ for $I_c$. One more observation is that the
relative width $\sigma/\langle k\rangle$ of the distributions
increases by a factor of about 3 in the evolution for different
initial average degree $\langle k\rangle$. Because of the constant
factor in the change of relative width in the evolution, the width
for the stationary state can be smaller or larger than that in the
initial state, depending on the value of initial average degree.

Another feature with the evolution of the network is the emergence
of isolated nodes. When the only one link a node has is a
connection to the node with highest degree under attack, the node
may have no connection to the network and becomes isolated after
reconstruction. Needless to say, the number of isolated nodes
depends on the evolution stage and the value of initial average
degree $\langle k \rangle$. The larger the initial average degree
$\langle k \rangle$, the less the number of isolated nodes at
fixed steps since the evolution. Our simulation shows that the
probabilities for a node to be isolated in the stationary state
are 0.03$\%$ and 25$\%$ with initial degree $\langle k \rangle =$
50 and 15, correspondingly. When initial mean degree $\langle k
\rangle$ is small, many nodes are isolated, and the isolation of
many nodes makes the nodes more independent and the system more
stable.

An important property of a network is its clustering coefficient.
To know the behavior of the clustering coefficient in the
evolution, we need to calculate the clustering coefficient of the
network in the initial state and in the stationary state. Let us
focus first on a selected node $i$ of the network. Suppose the
node have $k_{i}$ edges connecting its so-called nearest
neighbors. The maximum possible edges among $k_{i}$ nearest
neighbors is $k_{i}(k_{i}-1)/2$. We use $n_{i}$ to denote the
number of edges that actually exist among those neighbors. The
clustering coefficient of node $i$ is defined as:
\begin{equation}
C_{i}=\frac {n_{i}} {k_{i}(k_{i}-1)/2}.
\end{equation}

The clustering coefficient of the entire network is defined as the
average over the whole network,
\begin{equation}
C=\frac {1}{N}\sum_{i}C_{i}.
\end{equation}

We find that the clustering coefficient in the stationary state is
much larger than that in the initial state. For a network with
$N=10000$ nodes, the clustering coefficients in the initial state
are equal to 0.005 and 0.015 when the initial mean degrees
$\langle k\rangle$ are 50 and 150, respectively, while the
clustering coefficients in the stationary state are 0.23 and 0.1,
respectively.

From the fact that the network in the stationary state has the
large clustering coefficients together with some isolated nodes
and low average degrees, one can conclude that the system is
driven in its evolution to a state composed of quite a few highly
clustered small clusters. This is in sharp contrast with the
network in the initial state when the system has a small
clustering coefficient, almost no isolated nodes, and a high
average degree.

\begin{figure}
\includegraphics{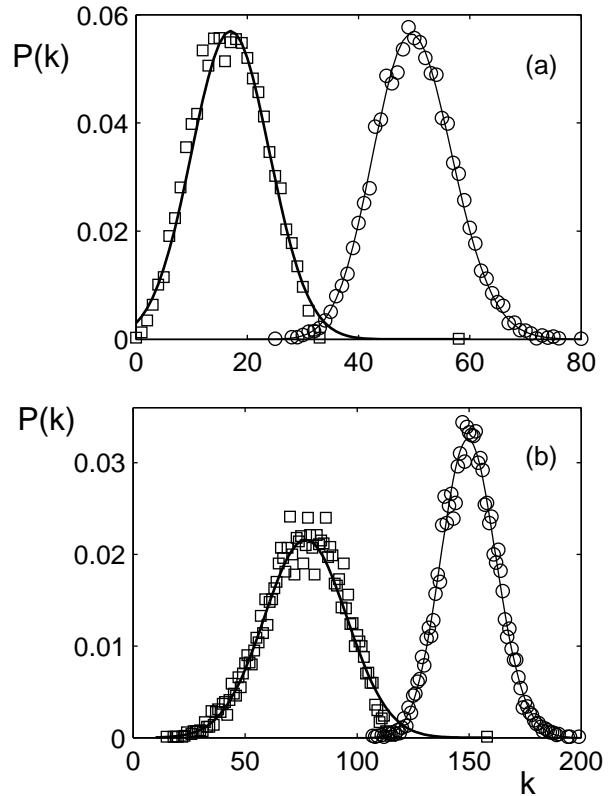}
\caption{\label{fig5}A comparison of degree distributions of the
network in the initial state (circles) with that in the stationary
state (squares) for (a) $N=10000$ and $p=0.005$; (b) $N=10000$ and
$p=0.015$. Solid lines are Gaussian fits.}
\end{figure}

\section{ Discussions and Conclusions}

In this paper, we study the evolution of random networks under
continuous attacks and subsequent reconstructions. We introduce a
new quality, invulnerability $I(s)$, to describe the safety of the
network. A stationary state with fixed $I(s)$ is observed during
the evolution of the network. The stationary value of
invulnerability $I_{c}$ is found to be independent of the network
size $N$ and the probability $p_{re}=p$ when the initial average
degree is fixed. $I_{c}$ shows a power-law dependence on the
initial mean degree $\langle k \rangle$, with the exponent is
about $-1.485$.

We give further information on the evolution of the properties of
the network. The first property is the evolution of the degree
distribution. The degree distributions of a network in both the
initial and the stationary states are found to be normal
distributions. After evolution, the peak position in the
distribution shifts to lower connection degree while the relative
width $\sigma/\langle k\rangle$ increases considerably. In the
stationary state, the edges and degrees in the whole network
decrease a lot and quite a few isolated nodes appear. The second
is the clustering coefficient. The clustering coefficient of a
network in the stationary state is much larger than that in the
initial state.

In summary, the stability of the network is found to be related
closely to the initial average degree $\langle k \rangle$ and has
little correlation with the network size $N$ and probability
$p_{re}=p$ separately. The stability of the network will increase
rapidly with the decrease of initial average degree $\langle k
\rangle$. The reason is that when $\langle k \rangle$ is small,
the edges and degrees decrease a lot and more isolated nodes
appear, which weaken the ability of nodes to communicate with each
other and make the network more stable. From the isolated nodes
and high clustering coefficient, we conclude that the network in
the stationary state is composed of some highly clustered small
clusters.

Still, there are many issues to be addressed, such as the
correlations and the fluctuations in $k_{max}$ in the evolution,
especially after the stationary state. These fluctuations may tell
us more the nature of the stationary state. The behavior of the
average degree in the evolution also may shed some light on the
dynamics of network. In addition, it is worthwhile to investigate
whether the evolution of random networks demonstrate
self-organized criticality (SOC), according to the similarity
between the evolution of invulnerability $I(s)$ with that of the
envelope function $G(s)$ in Bak-Sneppen model. All these topics
can not be covered in this paper, and will be discussed later.

\section*{Acknowledgments}

This work was supported in part by the National Natural Science
Foundation of China under grant No. 70271067 and by the Ministry
of Education of China under grant No. 03113.

\end{document}